\documentclass[a4paper]{jpconf}

\usepackage{amsmath,amssymb}
\usepackage{graphicx}
\usepackage{natbib}

\usepackage{epstopdf}

\begin{document}
\title{Cross-helicity effects and turbulent transport in magnetohydrodynamic flow}

\author{Nobumitsu Yokoi$^{1,}$\footnote[2]{Guest Researcher at the National Astronomical Observatory of Japan} and Guillaume Balarac$^3$}

\address{$^1$ Institute of Industrial Science, University of Tokyo, 4-6-1, Komaba, Meguro, Tokyo 153-8505, Japan}
\address{$^3$ Laboratoire des Ecoulements G\'{e}ophysiques et Industriels, University of Grenoble, France}

\ead{nobyokoi@iis.u-tokyo.ac.jp}

\begin{abstract}
In the presence of large-scale vortical motions and/or magnetic-field strains, the turbulent cross helicity (velocity--magnetic-field correlation in fluctuations) may contribute to the turbulent electromotive force and the Reynolds stress. These effects of cross helicity are considered to balance the primary effects of turbulence such as the turbulent magnetic diffusivity in magnetic-field evolution and the eddy viscosity in the momentum transport. The cross-helicity effects may suppress the enhanced transports due to turbulence. Physical interpretation of the effects is presented with special emphasis on the difference between the cross-helicity effect and the usual $\alpha$ or helicity effect in the dynamo action. The relative importance of the cross-helicity effect in dynamo action is validated with the aid of a direct numerical simulation (DNS) of the Kolmogorov flow with an imposed magnetic field. Several mechanisms that provide turbulence with the cross helicity are also discussed.
\end{abstract}

\section{\label{sec:level1}Introduction}
Cross helicity (velocity--magnetic-field correlation) is a quantity of fundamental importance at high magnetic Reynolds number magnetohydrodynamic (MHD) flow. The total cross helicity, as well as the MHD energy, is an inviscid invariant of the MHD system of equations, but a pseudoscalar which represents breakage of mirror-symmetry. Cross helicity is expected to play a certain role in the turbulent dynamo \citep[see] [and works cited therein]{Yoshizawa_etal_2004}. 

\begin{figure}[htb]
\begin{center}
    \includegraphics[width = 0.5\textwidth] {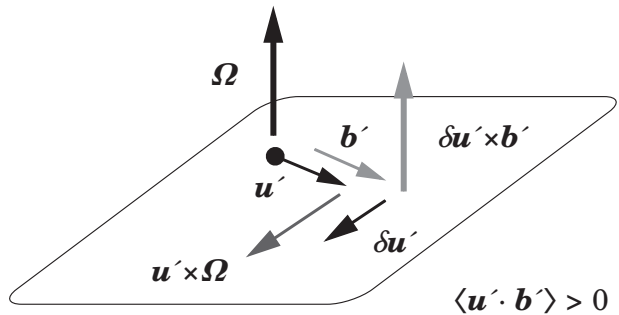}
    \caption{\label{fig:cross-helicity_effect}Physical interpretation of the cross-helicity effect. Redrawn from \citet{Yokoi_1999}.}
\end{center}
\end{figure}

	In the presence of a large-scale vortical motion, the turbulent cross helicity $\langle {{\boldsymbol{u}}' \cdot {\boldsymbol{b}}'} \rangle$ contributes to the electromotive force aligned with the large-scale vorticity ${\boldsymbol{\it\Omega}} (= \nabla \times {\boldsymbol{U}})$ (${\boldsymbol{u}}'$: velocity fluctuation, ${\boldsymbol{b}}'$: magnetic-field fluctuation, ${\boldsymbol{U}}$: mean velocity, $\langle{\cdots}\rangle$: ensemble average). A fluid element in the large-scale vorticity is subject to the Coriolis-like force ${\boldsymbol{u}}' \times {\boldsymbol{\it\Omega}}$. Provided that there is a positive (or negative) cross correlation between the velocity and magnetic-field fluctuations, a contribution to the electromotive force parallel (or antiparallel) to the mean vorticity, $\langle {\delta{\boldsymbol{u}}' \times {\boldsymbol{b}}'} \rangle \propto \langle {{\boldsymbol{u}}' \cdot {\boldsymbol{b}}'} \rangle {\boldsymbol{\it\Omega}}$, arises (figure~\ref{fig:cross-helicity_effect}). This is in marked contrast with a positive (or negative) turbulent helicity $\langle {{\boldsymbol{u}}' \cdot {\boldsymbol{\it\omega}}'} \rangle$, correlation between the velocity and vorticity fluctuations may contribute to the electromotive force antiparallel (or parallel) to the mean magnetic field [${\boldsymbol{\it\omega}}' (= \nabla \times {\boldsymbol{u}}')$: vorticity fluctuation]. 
	
	If we take the cross-helicity effects into account, the turbulent electromotive force ${\boldsymbol{E}}_{\rm{M}}$ is expressed as \citep{Yoshizawa_1990}
\begin{equation}
	{\boldsymbol{E}}_{\rm{M}}
	\equiv \langle {{\boldsymbol{u}}' \times {\boldsymbol{b}}'} \rangle
	= \alpha {\boldsymbol{B}}
	- \beta \nabla \times {\boldsymbol{B}}
	+ \gamma {\boldsymbol{\it\Omega}},
	\label{eq:turb_emf}
\end{equation}
where $\alpha$, $\beta$, and $\gamma$ are the transport coefficients expressed in terms of the residual helicity $\langle { - {\boldsymbol{u}}' \cdot {\boldsymbol{\it\omega}}' + {\boldsymbol{b}}'\cdot {\boldsymbol{j}}' } \rangle (\equiv H)$, the turbulent MHD energy $\langle {{\boldsymbol{u}}'{}^2 + {\boldsymbol{b}}'{}^2} \rangle/2 (\equiv K)$, and the cross helicity $\langle {{\boldsymbol{u}}' \cdot {\boldsymbol{b}}' } \rangle (\equiv W)$, respectively. 

	If we substitute (\ref{eq:turb_emf}) into the mean magnetic induction equation, we obtain
\begin{equation}
	\frac{\partial {\boldsymbol{B}}}{\partial t}
	= \nabla \times \left( {
		{\boldsymbol{U}} \times {\boldsymbol{B}}
	} \right)
	+ \nabla \times \left( {
		\alpha {\boldsymbol{B}}
		+ \gamma {\boldsymbol{\it\Omega}}
	} \right)
	- \nabla \times \left[ {
		(\eta + \beta) \nabla \times {\boldsymbol{B}}
	} \right]
	\label{eq:mean_mag_ind_eq}
\end{equation}
($\eta$: magnetic diffusivity). First, in the presence of turbulence, the effective magnetic diffusivity or resistivity is enhanced by turbulence ($\eta \rightarrow \eta + \beta$) with spatiotemporal variation of $\beta$. So, the coefficient $\beta$ is called the turbulent magnetic diffusivity or anomalous resistivity. Secondly, the $\alpha$- and $\gamma$-related terms express possible magnetic-field generation mechanisms due to pseudoscalars. If turbulence has some asymmetric properties represented by these pseudoscalars, the enhanced transport due to $\beta$ can be suppressed by these pseudoscalar effects. This suppression of turbulent transport is based on the dynamic balance between two turbulent effects: enhancement and suppression of the transport. This state is entirely different from the one subject only to the molecular transport. Turbulent itself is very strong, but due to the additional asymmetry represented by pseudoscalars, effective turbulent transport is dynamically suppressed. Then important point to see is how and how much such pseudoscalars can be present in turbulence. 

	In the long history of the dynamo study, special attention has been paid to the $\alpha$ or helicity effect [the first term in (\ref{eq:turb_emf})]. In contrast, the $\gamma$ or cross-helicity effect [the third term in (\ref{eq:turb_emf})] has been almost missing except for some limited number of works \citep[e.g.,][and works cited therein]{Yoshizawa_etal_2004}. One of the reasons of this missing may be attributed to the current treatment of turbulence; in most works, the large-scale velocity is neglected in the consideration of turbulence because of the Galilean invariance of the momentum equation. However, such treatment inevitably leads to the neglect of the large-scale shear effects. On the contrary, a large-scale rotation is ubiquitous and sometimes essential ingredient in astro/geophysical phenomena. In this sense, we should take effects of large-scale inhomogeneous flows into consideration of the realistic turbulent dynamo studies.
	
	The number of works that contain numerical simulations examining or validating the cross-helicity effect is very limited \citep{Hamba_1992,Nishino_Yokoi_1998,Yokoi_Hamba_2007,Yokoi_etal_2008,Sur_Brandenburg_2009}. In the present work, we examine the effect of turbulent cross helicity, and see the validity of the notion of cross-helicity dynamo. This must be a very interesting contribution to the astro/geophysical dynamo studies from the shear turbulence study.

	Organization of this paper is as follows. After an outlined presentation of the turbulent cross-helicity generation mechanisms with the aid of the evolution equation of the turbulent cross helicity in section~\ref{sec:level2}, a brief description of the MHD Kolmogorov flow is given in section~\ref{sec:level3}. Some of the basic numerical results are presented in section~\ref{sec:level4}, which include the analysis of the turbulent electromotive force (section~\ref{sec:level4-1}), the comparison of the induced fields with some approximate dynamo solutions (section~\ref{sec:level4-2}), the comparison of the spatial distributions of the turbulent cross helicity and its production rate (section~\ref{sec:level4-3}), and the turbulent cross helicity scaled by the turbulent MHD energy (section~\ref{sec:level4-4}). Concluding remarks with directions in the future work are presented in section~\ref{sec:level5}.

\section{\label{sec:level2}Cross-helicity generation mechanisms}
From equations of the velocity and magnetic-field fluctuations, ${\boldsymbol{u}}'$ and ${\boldsymbol{b}}'$, we obtain the equation for the turbulent cross helicity
\begin{equation}
	W = \langle {{\boldsymbol{u}}' \cdot {\boldsymbol{b}}'} \rangle
	\label{eq:W_def}
\end{equation}
as
\begin{equation}
	\frac{DW}{Dt} 
	\equiv \left( {
		\frac{\partial}{\partial t} + {\boldsymbol{U}}\cdot \nabla
	} \right) W
	= P_W - \varepsilon_W + T_W.
	\label{eq:W_eq}
\end{equation}
Here, $P_W$, $\varepsilon_W$, and $T_W$ are the production, dissipation, and transport rates of $W$. They are defined by
\begin{equation}
	P_W = - {\cal{R}}^{ab} \frac{\partial B^b}{\partial x^a} 
	- {\boldsymbol{E}}_{\rm{M}} \cdot {\boldsymbol{\it\Omega}},
	\label{eq:P_W_def}
\end{equation}
\begin{equation}
	\varepsilon_W = \left( {\nu + \eta} \right) \left\langle {
		\frac{\partial u'{}^b}{\partial x^a}
		\frac{\partial b'{}^b}{\partial x^a}
	} \right\rangle,
	\label{eq:eps_W_def}
\end{equation}
\begin{equation}
	T_W = {\boldsymbol{B}} \cdot \nabla K + \nabla \cdot {\boldsymbol{T}}'_W,
	\label{eq:T_W_def}
\end{equation}
where $\nabla \cdot {\boldsymbol{T}}'_W$ is the transport rate of $W$ that can be written in the divergence form. It comes from the higher order terms of ${\boldsymbol{u}}'$ and ${\boldsymbol{b}}'$, whose details are suppressed here. In (\ref{eq:P_W_def}), $\mbox{\boldmath$\mathcal{R}$}$ and ${\boldsymbol{E}}_{\rm{M}}$ are the Reynolds stress (combination of the Reynolds and turbulent Maxwell stresses in the MHD case) and the turbulent electromotive force, respectively. They are defined by
\begin{equation}
	{\mathcal{R}}^{\alpha\beta}
	= \left\langle {
		u'{}^\alpha u'{}^\beta - b'{}^\alpha b'{}^\beta
	} \right\rangle.
	\label{eq:Re_strss_def}
\end{equation}
and (\ref{eq:turb_emf}). The Reynolds stress is known to be expressed as
\begin{equation}
	{\mathcal{R}}^{\alpha\beta}
	= \frac{2}{3} K_{\rm{R}} \delta^{\alpha\beta}
	- \nu_{\rm{K}} {\mathcal{S}}^{\alpha\beta}
	+ \nu_{\rm{M}} {\mathcal{M}}^{\alpha\beta},
	\label{eq:Re_strss_exp}
\end{equation}
where $K_{\rm{R}} (\equiv \langle {{\boldsymbol{u}}'{}^2 - {\boldsymbol{b}}'{}^2} \rangle /2)$ is the turbulent MHD residual energy, and $\nu_{\rm{K}} [= (7/5) \beta]$ and $\nu_{\rm{M}} [= (7/5) \gamma]$ are the turbulent transport coefficients. The strain rates of the mean velocity and magnetic field, $\mbox{\boldmath${\mathcal{S}}$}$ and $\mbox{\boldmath${\mathcal{M}}$}$, are defined by
\begin{equation}
	{\mathcal{S}}^{\alpha\beta}
	= \frac{\partial U^\alpha}{\partial x^\beta}
	+ \frac{\partial U^\beta}{\partial x^\alpha},
	\label{eq:S_def}
\end{equation}
\begin{equation}
	{\mathcal{M}}^{\alpha\beta}
	= \frac{\partial B^\alpha}{\partial x^\beta}
	+ \frac{\partial B^\beta}{\partial x^\alpha}.
	\label{eq:M_def}
\end{equation}

	If we substitute the expressions for $\mbox{\boldmath$\mathcal R$}$ and ${\boldsymbol{E}}_{\rm{M}}$,  (\ref{eq:Re_strss_exp}) and (\ref{eq:turb_emf}), into the production rate of the turbulent cross helicity, $P_W$, (\ref{eq:P_W_def}), we obtain
\begin{equation}
	P_W 
	= \frac{1}{2} \nu_K {\mbox{\boldmath$\cal{S}$}}:{\mbox{\boldmath$\cal{M}$}}
	- \frac{1}{2} \nu_M {\mbox{\boldmath$\cal{M}$}}^2
	- \alpha {\boldsymbol{B}} \cdot {{\boldsymbol{\it\Omega}}}
	+ \beta {\boldsymbol{J}} \cdot {{\boldsymbol{\it\Omega}}}
	- \gamma {{\boldsymbol{\it\Omega}}}^2
	\label{eq:P_W_detail}
\end{equation}
[$\mbox{\boldmath$\cal{S}$}: \mbox{\boldmath$\cal{M}$} = {\mathcal{S}}^{ab} {\mathcal{M}}^{ab}$, $\mbox{\boldmath$\cal{M}$}^2 = {\mathcal{M}}^{ab} {\mathcal{M}}^{ab}$]. This shows that combinations of the mean-field shears coupled with the turbulent transport coefficients give the generation mechanisms of the turbulent cross helicity. For the detailed meaning of each term in the production, dissipation, and transport rates, and for the possible situations where the cross helicity is supplied to turbulence, the reader is referred to \citet {Yokoi_2011, Yokoi_Hoshino_2011}.

\section{\label{sec:level3}Magnetohydrodynamic Kolmogorov flow}
\begin{figure}[htb]
\begin{center}
    \includegraphics[width = 0.75\textwidth] {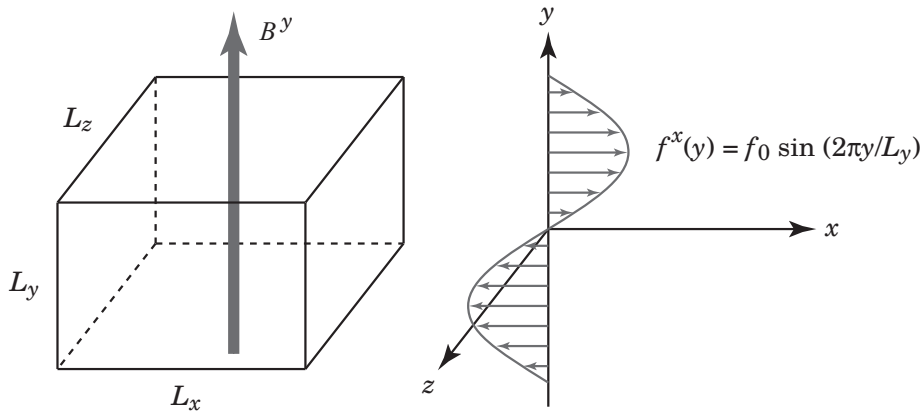}
    \caption{\label{fig:kolmogorov_flow}Kolmogorov flow with a magnetic field imposed.}
\end{center}
\end{figure}

Kolmogorov flow is a three-dimensional periodic flow with external forcing. This flow is homogeneous in the $x$ and $z$ directions, but due to the external forcing expressed as
\begin{equation}
	{\boldsymbol{f}} = \left( {f^x, f^y, f^z} \right) 
	= \left( {
		f_0 \sin \left( {{2\pi y}/{L^y}} \right), 0, 0
	} \right)
	\label{eq:forcing}
\end{equation}
($L^x$, $L^y$, $L^z$: box dimension), it is inhomogeneous in the $y$ direction (figure~\ref{fig:kolmogorov_flow}). The Kolmogorov flow provides a good test for investigating three-dimensional inhomogeneous turbulent flow simultaneously with the mean-velocity shear and the anisotropy \citep[][and works cited therein, also see Hamba, 1992]{Sarris_etal_2007}. In order to examine the basic properties of magnetohydrodynamic (MHD) turbulence, in addition to the external forcing, we impose a uniform large-scale magnetic field in the inhomogeneous ($y$) direction as
\begin{equation}
	{\boldsymbol{B}} = \left( {B^x, B^y, B^z} \right) = \left( {0, B_0, 0} \right).
	\label{eq:uniform_mag_fld}
\end{equation}

	In this work, with the aid of a direct numerical simulation of the Kolmogorov flow,  we address the question of whether the cross-helicity effect is relevant to the turbulent electromotive force ${\boldsymbol{E}}_{\rm{M}}$ or not. 
	
	It was reported that the statistics of turbulence depend on the several computational conditions in the Kolmogorov \citep{Sarris_etal_2007}. In this sense, to obtain quantitative results, we have to scrutinize these conditions, including the dependences on the Reynolds number, aspect ratio of the box, strength of the imposed magnetic field, etc. We also have to take care of the characteristic length scales in the three directions (in particular in the direction of the imposed magnetic field) in comparison with the domain sizes. However, here our attention is focused on the basic behaviours of the cross-helicity effect in ${\boldsymbol{E}}_{\rm{M}}$. So, we only consider a situation of a cubic box with the numerical discretization of $256^3$. In the following, the operator $\langle \cdots \rangle$ denotes averaging over the homogeneous ($x$ and $z$) directions and ensemble averaging over 70 independent realizations in time. The primed quantities correspond to deviations from the averages. Each realization shows large spatial variations. In order to get more smooth statistics, we have to take larger statistical ensembles.

\section{\label{sec:level4}Basic results and discussions}
The situation of Kolmogorov flow may be unrealistic in that it is difficult to experimentally generate such a periodic forcing (\ref{eq:forcing}) in an unbounded flow. Nevertheless, this flow gives a test for investigating three-dimensional turbulence with inhomogeneity, anisotropy, and velocity shear. So, with an imposed magnetic field, it gives a suitable situation to validate the cross-helicity effect in a magnetohydrodynamic turbulent flow. In this work, we examine mainly two points: (i) how the cross-helicity-related term contribute to the turbulent electromotive force; and (ii) how the cross helicity is supplied to turbulence.

\subsection{\label{sec:level4-1}Turbulent electromotive force}
\begin{figure}[htb]
\begin{center}
    \includegraphics[width = 0.5\textwidth] {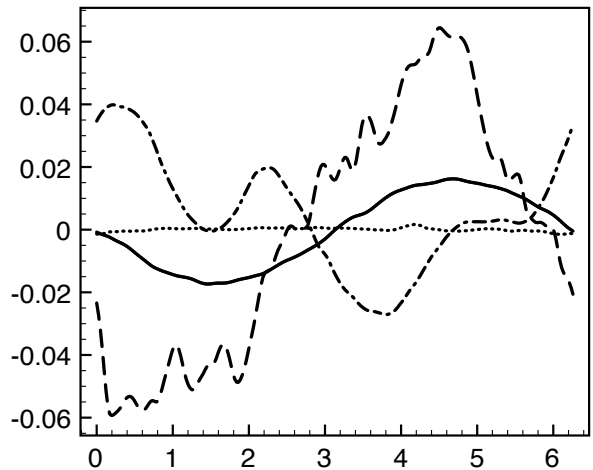}
   \caption{\label{fig:E_M_comparison}Spatial distribution of the turbulent electromotive force and each dynamo term in the Kolmogorov flow versus $y$: ------ $\langle{{\boldsymbol{u}}' \times {\boldsymbol{b}}'}\rangle^z$; $\cdots\cdots$ $\alpha B^z$; -- -- -- $-\beta J^z$;  -- $\cdot$ -- $\gamma \Omega^z$. }
\end{center}
\end{figure}

First we consider the turbulent electromotive force ${\boldsymbol{E}}_{\rm{M}} (\equiv \langle{ {\boldsymbol{u}}' \times {\boldsymbol{b}}' }\rangle)$. The spatial distribution of the $z$ component of  ${\boldsymbol{E}}_{\rm{M}}$ is shown in figure~\ref{fig:E_M_comparison}. The profile of ${\boldsymbol{E}}_{\rm{M}}$ is sinusoidal as expected from the external forcing. 

	In figure~\ref{fig:E_M_comparison}, we also showed the spatial distribution of each term of the right-hand side in (\ref{eq:turb_emf}). We see from this figure that the $\alpha$- or helicity-related term is much less than the $\beta$- or turbulent magnetic diffusivity-related term and $\gamma$- or cross-helicity-related terms. The main contribution to the turbulent electromotive force in this case is attained by the balance between the $\beta$- and $\gamma$-related terms: 
\begin{equation}
	{\boldsymbol{E}}_{\rm{M}} 
	\simeq - \beta {\boldsymbol{J}} 
	+ \gamma {\boldsymbol{\it\Omega}}.
	\label{eq:main_balance_E_M}
\end{equation}
Namely, the turbulent magnetic-diffusivity effect is balanced by the turbulent cross-helicity effect rather than by the turbulent helicity effect.

	This result certainly shows the importance of the cross-helicity effect as compared to the $\alpha$ or helicity effect in the dynamo action in a shear flow situation.

\subsection{\label{sec:level4-2}Dynamo solutions}
If the $\alpha$ or helicity effect is negligibly small in the turbulent electromotive force, the mean induction equation (\ref{eq:mean_mag_ind_eq}) is written as
\begin{equation}
	\frac{\partial {\boldsymbol{B}}}{\partial t}
	= \nabla \times \left[ {
		{\boldsymbol{U}} \times {\boldsymbol{B}}
		- \left( {\beta + \eta} \right) {\boldsymbol{J}}
		+ \gamma \mbox{\boldmath$\it\Omega$}
	} \right].
	\label{eq:cr_dyn_mean_induct_eq}
\end{equation}
In the present geometry of the Kolmogorov flow, this reads
\begin{equation}
	\frac{\partial B^x}{\partial t}
	= B_0 \frac{\partial U^x}{\partial y}
	- \frac{\partial}{\partial y} \left[ {\left( {\beta + \eta} \right) J^z} \right]
	+ \frac{\partial}{\partial y} \left( {\gamma \Omega^z} \right).
	\label{eq:B_x_ind_eq}
\end{equation}
The first term in (\ref{eq:B_x_ind_eq}) arises from the velocity inhomogeneity in the direction of the mean magnetic field, which corresponds to the so-called $\Omega$ effect.

\begin{figure}[htb]
\begin{center}
    \includegraphics[width = 1.\textwidth] {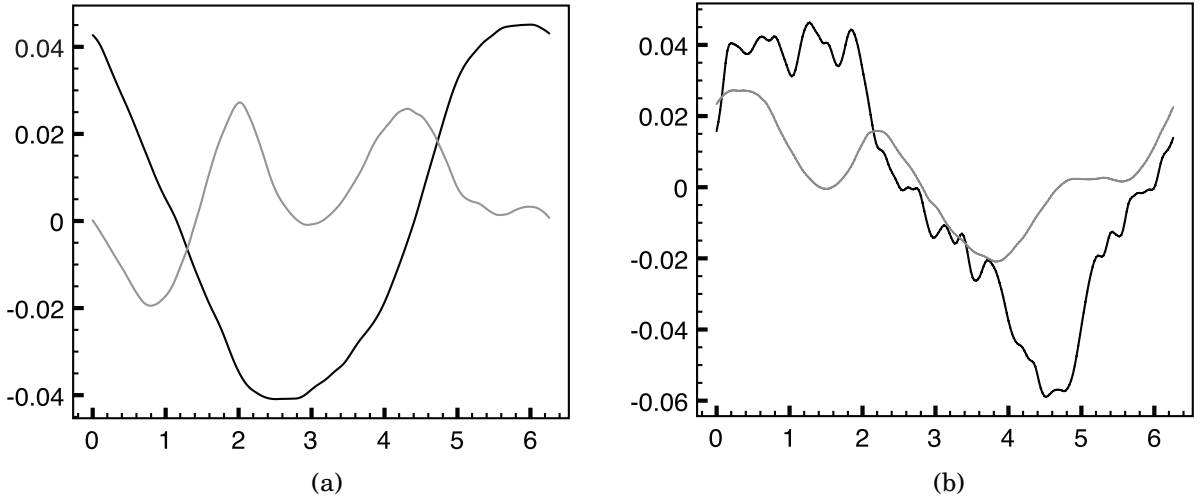}
   \caption{\label{fig:dyn_sol}Spatial distribution of the induced field and approximate dynamo solutions versus $y$. (a) Magnetic field $B^x$ and velocity $(W/K) U^x$: ------ (black) $B^x$; ------ (gray) $(W/K) U^x$. (b) Electric-current density $J^z$ and vorticity $(W/K) \Omega^z$: ------ (black) $J^z$; ------ (gray) $(W/K) \Omega^z$ versus $y$. Constant factor (model constant) $C$ is not included in the dynamo-solution expressions.}
   \end{center}
\end{figure}

	For a stationary state, we may consider an approximate dynamo solution for the induced field \citep{Yoshizawa_Yokoi_1993,Yokoi_Hoshino_2011} as
\begin{equation}
	\delta{\boldsymbol{B}} 
	= \frac{\gamma}{\beta} {\boldsymbol{U}}
	= C \frac{W}{K} {\boldsymbol{U}},
	\label{eq:dynamo_sol_B}
\end{equation}
where $C$ is the model constant of $O(10^{-1})$. Here we dropped the magnetic diffusivity $\eta$ as compared with the turbulent counterpart $\beta$. However, in case $\gamma / \beta$ or $W/K$ spatially varies, the approximate dynamo solution (\ref{eq:dynamo_sol_B}) is not the case. In figure~\ref{fig:dyn_sol}(a) we plot the induced magnetic field $B^x$ in comparison with the approximate dynamo solution $(W/K) U^x$.

	If the turbulent magnetic diffusivity is fully balanced by the cross-helicity effect in the turbulent electromotive force, the mean electric-current density is aligned with the mean vorticity as
\begin{equation}
	\delta{\boldsymbol{J}} 
	= \frac{\gamma}{\beta} \mbox{\boldmath$\it\Omega$}
	= C \frac{W}{K} \mbox{\boldmath$\it\Omega$}.
	\label{eq:dynamo_sol_J}
\end{equation}

	We plot the induced mean electric-current density $J^z (= - \partial B^x / \partial y)$ in comparison with the approximate dynamo solution $(W/K) \Omega^z [= - (W/K) (\partial U^x / \partial y)]$ in figure~\ref{fig:dyn_sol}(b). Unlike (\ref{eq:dynamo_sol_B}), the general tendency of $J^z$ is in agreement to some extent with the counterpart of $(W/K) \Omega^z$, although a further analysis of each term of (\ref{eq:B_x_ind_eq}) is needed.

\subsection{\label{sec:level4-3}Cross helicity and its production rate}
\begin{figure}[htb]
\begin{center}
    \includegraphics[width = 0.5\textwidth] {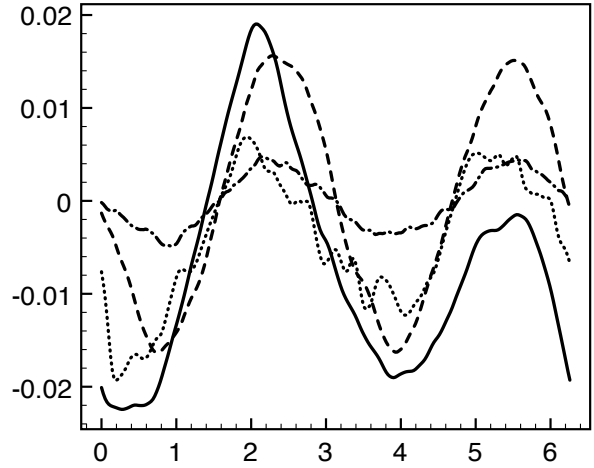}
   \caption{\label{fig:P_W_comparison}Spatial distribution of the turbulent cross helicity and its production-rate terms in the Kolmogorov flow versus $y$: ------ $W/\tau$; $\cdots\cdots$ $-{\cal{R}}^{ab} (\partial B^a / \partial x^b)$; -- -- -- $- {\boldsymbol{E}}_{\rm{M}} \cdot {\boldsymbol{\it\Omega}}$; -- $\cdot$ -- ${\boldsymbol{B}} \cdot \nabla K$.}
\end{center}
\end{figure}

Next, we consider the cross-helicity generation mechanisms in the Kolmogorov flow. For this purpose, we show in figure~\ref{fig:P_W_comparison} that the spatial distribution of the turbulent cross helicity scaled by the characteristic time of turbulence, $W/\tau$. As for the time scale, we adopt the energy cascade time defined by the turbulent MHD energy $K$ divided by its dissipation rate $\varepsilon$ as
\begin{equation}
	\tau = K / \varepsilon
	\label{eq:time_scale}
\end{equation}
with
\begin{equation}
	K = \left\langle {
		{\boldsymbol{u}}'{}^2 + {\boldsymbol{b}}'{}^2
	} \right\rangle /2,
	\label{eq:K_def}
\end{equation}
\begin{equation}
	\varepsilon = \nu \left\langle {
		\left( {\frac{\partial u'{}^b}{\partial x^a}} \right)^2
	} \right\rangle
	+ \eta \left\langle {
		\left( {\frac{\partial b'{}^b}{\partial x^a}} \right)^2
	} \right\rangle.
	\label{eq:eps_def}
\end{equation}
We also plot each term in the production rate of the turbulent cross helicity, (\ref{eq:P_W_def}) and (\ref{eq:T_W_def}), in figure~\ref{fig:P_W_comparison}.

	This figure shows that the spatial distribution of the turbulent cross helicity is in good agreement with the counterparts of the cross-helicity generation mechanisms. Of course we need to delve further into the detailed budget of the cross-helicity evolution (\ref{eq:W_eq}) including the dissipation and transport rates. With this reservation in mind, we see from this result that the production rate provides a good measure for the turbulent cross helicity itself. 

	Estimate of the turbulent cross helicity with its spatial distribution from observation is a challenging problem. All the three components of the velocity and magnetic-field fluctuations should be measured simultaneously. In actual observations, we often encounter a situation where the direct measurement of the cross helicity is impossible or hard to achieve. Even in such a case, if we can measure the production rate of cross helicity, we may estimate the turbulent cross helicity. In the detailed expression (\ref{eq:P_W_detail}) for the production rate of $W$, the second and fifth terms, $-\nu_{\rm{M}} \mbox{\boldmath$\mathcal{M}$}^2 / 2$ and $- \gamma \mbox{\boldmath$\it \Omega$}^2$, always work for the reduction of $W$. If we concentrate our attention on the genuine generation mechanisms of $W$, which come from the primary effects of turbulence such as the turbulent-energy- or $\beta$- and $\nu_{\rm{K}}$-related terms, then we have
\begin{equation}
	P_W 
	\simeq \frac{1}{2} \nu_{\rm{K}} 
		\mbox{\boldmath$\cal{S}$}:\mbox{\boldmath$\cal{M}$}
	+ \beta {\boldsymbol{J}} \cdot {\boldsymbol{\it\Omega}}.
	\label{eq:P_W_primary}
\end{equation}
In this case, without knowing any information on the turbulent cross helicity itself, we may be able to estimate $W$ by measuring the large-scale field configuration represented by the mean velocity and magnetic-field strains, $\mbox{\boldmath$\mathcal{S}$}$ and $\mbox{\boldmath$\mathcal{M}$}$, the mean electric current density ${\boldsymbol{J}}$, and the mean vorticy ${{\boldsymbol{\it\Omega}}}$, etc.

\subsection{\label{sec:level4-4}Scaled cross helicity}
As we see in (\ref{eq:dynamo_sol_J}), the proportional coefficient $\gamma/\beta$ or $W/K$ (the turbulent cross helicity $\langle {{\boldsymbol{u}}' \cdot {\boldsymbol{b}}'} \rangle$ scaled by the turbulent MHD energy $\langle {{\boldsymbol{u}}'{}^2 + {\boldsymbol{b}'{}^2}} \rangle /2$) combines the large-scale vorticity $\mbox{\boldmath$\it\Omega$}$ with the large-scale electric-current density ${\boldsymbol{J}}$ induced by the dynamo action. If the turbulent cross helicity is negligibly small compared to the turbulent MHD energy, we have substantially no dynamo effects due to the cross helicity. In this sense, $W/K$ is one of the most important quantities in the arguments of the cross-helicity effects. 

\begin{figure}[htb]
\begin{center}
    \includegraphics[width = 0.5\textwidth] {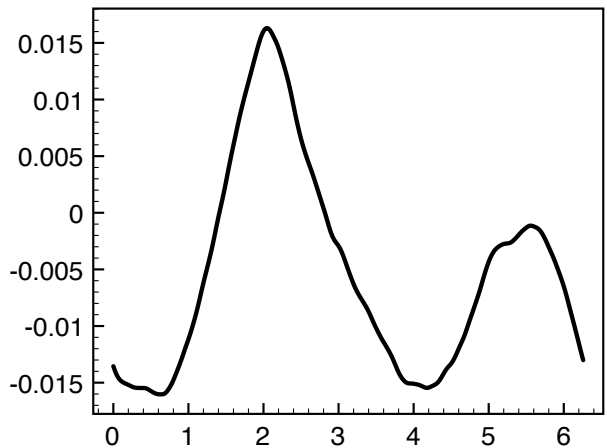}
   \caption{\label{fig:W_over_K}Spatial distribution of the turbulent cross helicity $W (= \langle {{\boldsymbol{u}}' \cdot {\boldsymbol{b}}'} \rangle)$ scaled by the turbulent MHD energy $K (= \langle {{\boldsymbol{u}}'{}^2 + {\boldsymbol{b}}'{}^2} \rangle/2)$ in the Kolmogorov flow versus $y$.}
\end{center}
\end{figure}

	The spatial distribution of the scaled cross helicity $W/K$ versus $y$ is plotted in figure~\ref{fig:W_over_K}.

	We see from this figure that the magnitude of the scaled cross helicity is $O(10^{-3})-O(10^{-2})$ in this case. It is worth noting here that the absolute value of $W/K$ is mathematically bounded as
\begin{equation}
	\frac{|W|}{K} 
	= \frac{|\langle {{\boldsymbol{u}}' \cdot {\boldsymbol{b}'}} \rangle|}
		{\langle {{\boldsymbol{u}}'{}^2 + {\boldsymbol{b}}'{}^2} \rangle /2}
	\le 1.
	\label{eq:W_over_K_bound}
\end{equation}
One self-evident but important consequence of this boundedness is that the turbulent cross helicity $W$ can not be sustained in the absence of the turbulent MHD energy ($K=0$).

	As the $W$ evolution equation (\ref{eq:W_eq}) shows, the value of $W/K$ must depend on the magnitude of the velocity shear and especially on the magnitude of the imposed magnetic field, $B_0$. To obtain more quantitative understanding, we have to delve into the detailed budgets of both the turbulent cross-helicity and MHD-energy evolutions. The highest value of $|W|/K$ is realized only when the velocity fluctuation ${\boldsymbol{u}}'$ is totally aligned with the magnetic-field fluctuation ${\boldsymbol{b}}'$ with equipartition of the turbulent kinetic and magnetic energies (Alfv\'{e}nic state):
\begin{equation}
	{\boldsymbol{u}}' = \pm {\boldsymbol{b}}'
	\label{eq:Alfven_state}
\end{equation}
This state is often observed in the solar-wind turbulence near the Sun because of the dominant effect of the Alfv\'{e}n waves. On the other hand, in the analysis of the galactic dynamos using the cross-helicity effect, it was inferred that $|W|/K$ of $O(10^{-3})-O(10^{-2})$ is large enough to explain the magnetic-field strength of several spiral galaxies \citep{Yokoi_1996}.

\section{\label{sec:level5}Concluding remarks}
On the contrary to the current understanding on the turbulent dynamo, where the $\alpha$ effect combined with the differential-rotation effect (called the $\Omega$ effect) plays the main role in the generation and evolution of the dynamo magnetic field, the relative importance of the cross-helicity effect to the $\alpha$ effect is suggested in the present work. This may pave a new way for the turbulent dynamo model for the astro/geophysical context. At least, the cross-helicity effect, in other words, the effect of the large-scale shear flow, should be taken into account in the dynamo study as a complementary effect to the well-known $\alpha$ or helicity effect.

	A large-scale magnetic field alters statistical properties of turbulence compared to the hydrodynamic turbulence. For the sake of further quantitative analysis, the dependences of the simulation results on the magnitudes of the imposed magnetic field, velocity shear, and on the grid size may be important.  These points should be further included in the forthcoming paper.

	As a final remark, we refer to the cross-helicity effect in the momentum equation. In the momentum equation, the turbulent transport is expressed by the Reynolds stress (\ref{eq:Re_strss_def}). Since the cross-helicity or $\nu_{\rm{M}}$-related term appears in the Reynolds-stress expression (\ref{eq:Re_strss_exp}) as a balancer with the the eddy viscosity $\nu_{\rm{K}}$, the cross helicity is expected to play an important role also in the suppression of turbulent linear- and angular-momentum transports. In addition, the cross-helicity effect may play important role in the momentum transport through the mean Lorentz force. The mean electric current ${\boldsymbol{J}}$ configuration induced by the cross-helicity effect is in general not aligned with the large-scale magnetic field (${\boldsymbol{J}} \nparallel {\boldsymbol{B}}$). This is in marked contrast with the counterpart by the $\alpha$ or helicity effect, where ${\boldsymbol{J}}$ is aligned with ${\boldsymbol{B}}$ (${\boldsymbol{J}} \times {\boldsymbol{B}} = 0$, force-free field). This feature of the cross-helicity effect was fully utilized in the studies of the inner oscillations in the Sun \citep{Itoh_etal_2005} and flow-turbulence interaction in magnetic reconnection \citep{Yokoi_Hoshino_2011}.
		
	In the context of solar physics, the differential rotation of the fluid inside the Sun has been left as an unsolved big problem. Recent elaborated numerical simulations showed that the Reynolds stress plays a key role in the angular-momentum transport inside the Sun \citep{Brun_Miesch_Toomre_2004, Miesch_Toomre_2009}. In order to address this problem, we should examine the Reynolds stress and its model expression (\ref{eq:Re_strss_exp}). In addition to the analysis of the turbulent electromotive force mentioned in section~\ref{sec:level4}, we will discuss the Reynolds stress expressions in the future work.

\ack
This work was started during the period both authors were staying at the Center for Turbulence Research (CTR), Stanford University and NASA Ames on the occasion of the CTR Summer Program 2010. They would like to express their thankfulness for the hospitality the CTR staffs extended to them. Part of this work was performed during the period one of the authors (NY) stayed at the Nordic Institute for Theoretical Physics (NORDITA) as a visiting researcher (February and March, 2011).

\bibliographystyle{jfm}
\bibliography{yokoi_balarac_etc13}

\end{document}